\documentclass[pdflatex,sn-mathphys-num]{sn-jnl}% Math and Physical Sciences Numbered Reference Style
\usepackage{graphicx}%
\usepackage{multirow}%
\usepackage{amsmath,amssymb,amsfonts}%
\usepackage{amsthm}%
\usepackage{mathrsfs}%
\usepackage[title]{appendix}%
\usepackage{xcolor}%
\usepackage{textcomp}%
\usepackage{manyfoot}%
\usepackage{booktabs}%
\usepackage{algorithm}%
\usepackage{algorithmicx}%
\usepackage{algpseudocode}%
\usepackage{listings}%

\usepackage{tikz}
\usepackage{caption}
\usepackage{listings}
\usepackage{xcolor}

\lstdefinelanguage{scasp}{
    morekeywords={not},
    sensitive=true,
    morecomment=[l]{\%},
    morestring=[b]"
}

\usetikzlibrary{shapes,arrows,positioning,fit,backgrounds,calc}

\tikzset{
    block/.style = {rectangle, draw, fill=blue!15, text width=6em, text centered, rounded corners, minimum width=2.5cm, minimum height=2cm, thick, inner sep=3pt, font=\small\bfseries},
    simblock/.style = {rectangle, draw, fill=green!15, text width=6em, text centered, rounded corners, minimum width=2.5cm, minimum height=2cm, thick, inner sep=3pt, font=\small\bfseries},
    serverblock/.style = {rectangle, draw, fill=red!15, text width=6em, text centered, rounded corners, minimum width=2.5cm, minimum height=2cm, thick, inner sep=3pt, font=\small\bfseries},
    arrow/.style = {thick, ->, >=stealth},
    darrow/.style = {thick, <->, >=stealth},
    label/.style = {font=\small, align=center},
    title/.style = {font=\small\bfseries, align=center}
}

\theoremstyle{thmstyleone}%
\theoremstyle{thmstyletwo}%
\theoremstyle{thmstylethree}%
\begin{document}

\title[Article Title]{Complex Autonomous UAV Task Execution and Decision-Making With s(CASP)}

\author[1]{\fnm{Keegan} \sur{Kimbrell}}\email{keegan.kimbrell@utdallas.edu}

\author[1]{\fnm{Alexis R.} \sur{Tudor}}\email{alexisrenee1@gmail.com}

\author[1]{\fnm{Peter Van} \sur{Vu}}\email{peter.vu@utdallas.edu}

\author[2]{\fnm{Trevor} \sur{Bihl}}\email{bihlt@ohio.edu}

\author[3]{\fnm{Doug} \sur{Slattery}}\email{doug@sybor-tech.com}

\author[1]{\fnm{Gopal} \sur{Gupta}}\email{gupta@utdallas.edu}

\affil[1]{\orgdiv{University of Texas at Dallas}, \orgname{ALPS Lab}, \orgaddress{\street{890 Franklyn Jenifer Dr}, \city{Richardson},  \state{Texas}, \postcode{75080}, \country{USA}}}

\affil[2]{\orgdiv{Ohio University}, \orgname{Electrical Engineering and Computer Science}, \orgaddress{\street{Stocker Center}, \city{Athens},  \state{OH}, \postcode{45701}, \country{USA}}}

\affil[3]{\orgname{Sybor-Tech}, \city{McKinney}, \state{TX}, \postcode{75072},  \country{USA}}

%%==================================%%

\abstract{Autonomous unmanned aerial vehicles (UAVs) must operate safely in dynamic environments and adapt to changing mission conditions. Although deep learning approaches have shown strong performance for navigation and perception, they are often difficult to explain, verify, and modify for safety-critical tasks. We propose a symbolic state-centered UAV agent using the s(CASP) answer set programming system, enabling autonomous task execution with constraint-based commonsense reasoning in a high-fidelity Unreal Engine 5 environment.
We fully implement prior work on the VECSR-A system to support multi-step autonomous behaviors including navigation, search, debris detection, precision spraying, object transport, and inspection. The UAV reasons over environmental and spatial constraints, dynamically revising plans when tasks fail or data is insufficient. Because decisions are based on commonsense reasoning, they are guaranteed to be correct and explainable. 
We evaluate the feasibility of s(CASP) for UAV control in realistic simulated missions. Results show that our framework enables explainable, adaptive autonomy without retraining, handling complex constraint-aware decisions and dynamic task reevaluation.}


%%==================================%%

\maketitle
\section{Introduction}\label{sec1}

The deployment of autonomous unmanned aerial vehicles (UAVs) in real-world environments presents fundamental challenges that extend beyond traditional path planning. While adaptive UAV planning and autonomous task generation capabilities have been demonstrated with hierarchical planning and probabilistic decision logic \cite{bib6}, the lack of declarative symbolic reasoning resulted in limited explainability. Although recent work has demonstrated the feasibility of symbolic reasoning for UAV navigation and collision avoidance using the VECSR-A framework \cite{bib1}, autonomous missions require a range of adaptive behaviors. UAVs operating in disaster response, inspection, or agriculture must navigate their environment safely and plan and execute multi-step tasks in real time.

Deep learning approaches, despite their strong empirical performance in perception and control tasks, suffer from limitations in reliability and safety \cite{bib8}. Specifically, machine learning solutions are "black boxes" that produce unreliable decisions that resist human interpretation and verification \cite{bib8}. These models require extensive offline training, such as training a neural network using thousands of hours of video \cite{bib9}, and cannot be easily modified once deployed. For UAVs operating in communications-limited environments where retraining is infeasible and reliability is critical, the inability to explain or adjust autonomous decisions undermines trust and safety in meeting mission requirements \cite{bib14}.

We address these challenges by fully implementing the proposed VECSR-A architecture with commonsense reasoning supported by s(CASP). We use this complete implementation to support multi-step task execution in a simulated Unreal Engine 5 city environment using Project AirSim \cite{bib2}. Our UAV agent uses s(CASP), a top-down, goal-directed, constraint answer set programming (ASP) language, to perform commonsense reasoning on environmental and spatial constraints \cite{bib3}. VECSR-A links the symbolic knowledge behind drone autonomy to the simulated environment. We reason over a dynamically updated knowledge base to achieve higher-level autonomous behaviors in the agent through hierarchical task decomposition.

We demonstrate this across three high-level capabilities: (i) autonomous navigation and search, (ii) environment interaction, and (iii) constraints and adaptation. We demonstrate these capabilities to complete five core tasks: (i) object inspection, (ii) debris detection, (iii) trespassers, (iv) spraying, and (v) transport/delivery. Further, we explain how the s(CASP) system and symbolic decision-making guarantee explainable, justifiable, and correct actions for each task and provide an empirical evaluation of the framework through mean execution time per task alongside ablation studies that isolate the contributions of its knowledge database. Our results demonstrate that this framework provides a flexible approach to explainable autonomy capable of handling dynamic tasks.

The main contributions of this paper are: (1) fully implementing VECSR-A to support multi-step autonomous task execution via s(CASP) goal-directed reasoning in the high-fidelity Project AirSim; (2) demonstrating the benefits that symbolic commonsense reasoning and task decomposition has over deep learning models for drone control; and (3) validating the framework empirically through runtime analysis across mission scenarios, confirming real-time decision-making in dynamic urban environments.

The remainder of the paper is structured as follows: Section 2 covers background on Unreal Engine 5, Project AirSim, s(CASP), and VECSR-A. Section 3 describes the system architecture and task decomposition. Section 4 outlines the experimental setup and evaluation methodology. Section 5 reports runtime performance and ablation results. Finally, Sections 6 and 7 discuss implications, related work, future directions, and summarize contributions.

\section{Background}\label{sec2}

\subsection{s(CASP) Goal-Directed Answer Set Programming}

~~~~ s(CASP) is a query-driven, goal-directed implementation of ASP that supports constraints. Using s(CASP) eliminates the need for full grounding in drone navigation tasks. Unlike traditional ASP, which grounds the entire program and transforms it before solving it with a propositional SAT solver, s(CASP) performs a top-down, goal-directed search without grounding \cite{bib3}. This allows the agent to handle complex reasoning problems that would otherwise be infeasible to ground, while additionally producing a proof tree that serves as a verifiable justification for each action.

Answer set programming (ASP), as formalized by \citet{bib10} supports nonmonotonic reasoning based on stable model semantics. This provides key advantages over traditional logic programming paradigms. However, most ASP solvers require grounding to find answer sets, including the most popular solver, \texttt{clingo} \cite{bib4}. Grounding is a process whereby all variables in an answer set program are instantiated with all possible constants that appear in the program, prior to solving the program using a SAT solver or similar. However, grounding the program results in a combinatorial explosion, and much of the research in grounding-based ASP centers on optimizing the grounding process \cite{bib12}. Therefore, for our realistic scenarios for reasoning, we instead use goal-directed ASP supported in the s(CASP) system \cite{bib3}. 

Rather than necessarily providing full answer sets, s(CASP) computes partial answer sets that satisfy a query. This enables us to use default negation (negation-as-failure), strong negation (or classical negation), as well as abducibles and constraints through even and odd loops over negation. For example, if we had a location that we want to assume is safe in the absence of concrete information that it is not, such as a docking location, we could use the following abducible:

\begin{lstlisting}
safe(dock) :- not unsafe(dock).
unsafe(dock) :- not safe(dock).
\end{lstlisting}

Thus, in the absence of concrete evidence the dock is unsafe, we generate multiple worlds where the dock is safe or unsafe depending on the rest of the program. Conversely, to disallow certain states, we add constraints. As an example, say we do not want to allow any answer set that entails an unsafe location for the drone, where an unsafe location \texttt{X} is defined as 
\lstinline{unsafe(X) :- on_fire(X), no_fly_zone(X).}, we would add the following rule to our knowledge base:

\begin{lstlisting} 
p :- drone_location(X), unsafe(X), not p. 

\end{lstlisting}

This means that if an answer set will contain \lstinline{drone_location(X)} and \lstinline{unsafe(X)}, that world will be pruned and such an answer set would be barred from existing. This is shortened in s(CASP) to \lstinline{:- drone_location(X), unsafe(X).} for convenience. These logical techniques and others allow for the representation of complex world state and human-like commonsense reasoning. 

A further advantage of s(CASP) is its enhanced explainability. The s(CASP) system offers transparency like other logic programming paradigms, but also provides full justifiability through backward-chaining execution from a query. Every query is therefore supported by its component reasoning steps in a format readable by both humans and machines. This justifiability is essential for mission-critical operations and for correcting the system in the field without costly retraining.

\subsection{VECSR}

\label{sec:vecsr}

To deploy s(CASP) in a realistic setting, it must be integrated with a real-world environment. We build upon the Virtually-Embodied Common Sense Reasoner (VECSR) \cite{bib5}, a framework that connects s(CASP) to both simulated and physical environments. The key advantage of VECSR is that it provides a foundational link between a simulated world and s(CASP)'s commonsense reasoning capabilities, enabling construction of a world model that is explainable and justifiable from an environment. 

Another benefit corresponds to a problem shared by most logic-based approaches: execution time. VECSR provides several static program optimizations to improve the processing speed of s(CASP) reasoning by reducing the size of the relevant program. This approach utilizes a tenet of the SOAR architecture \cite{bib13} wherein only the information that is needed is kept readily accessible in short-term memory to maintain efficiency. 

\begin{figure*}[h]

\centering

\includegraphics[width = 300px]{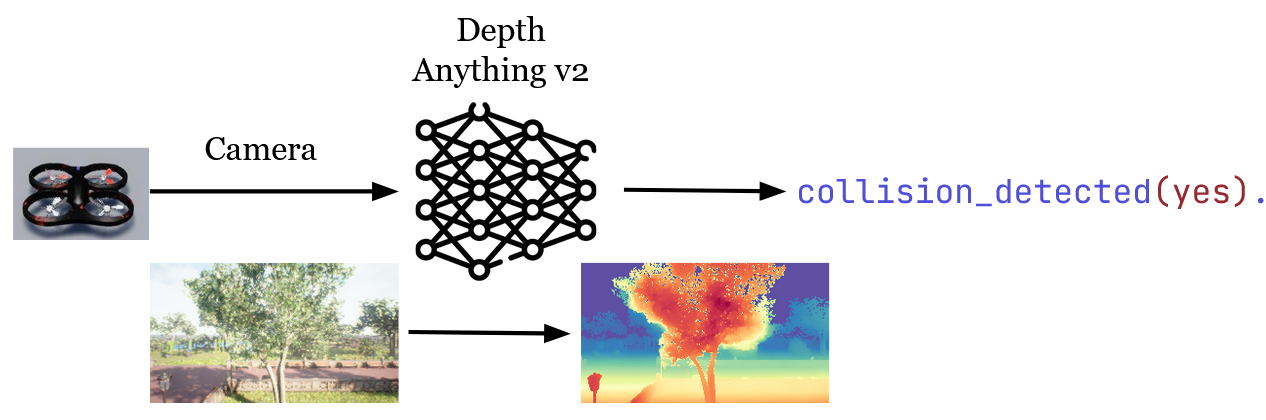}

\caption{The conversion process by VECSR-A module from image to collision detected fact.}

\label{fig:vecsra}

\end{figure*}

The VECSR system has been adapted for use in drone autonomy, resulting in the preliminary VECSR-A system \cite{bib1}. VECSR-A developed  a basic theoretical design for using VECSR with Microsoft’s AirSim \cite{bib11} for drone autonomy. The proposed architecture saw VECSR-A acting as an orchestrator using s(CASP) for modules that perform different types of analysis on the raw data that would come in from a drone. As an example proof of concept, an image model was used with data from the onboard camera to assess depth and detect collisions. That information was then input into s(CASP) facts for reasoning with the knowledge base, as in Figure \ref{fig:vecsra}. 

The basic idea is to use machine learning-based systems to translate the sensory inputs from the environment (e.g., scene captured through the camera) into logic predicates that encapsulate a description of the surroundings. Commonsense knowledge encoded in s(CASP), along with these captured logic predicates are then used to make navigation decisions. The idea is to navigate the drone autonomously as if a human was sitting on it and piloting it using their sensing and commonsense reasoning capabilities.

\subsection{From AirSim to Project AirSim}

The proof-of-concept VECSR-A system referenced in Section \ref{sec:vecsr} originally utilized Microsoft AirSim \cite{bib2} which was widely regarded as a leading high-fidelity simulation platform for autonomous drone research and development \cite{bib11}. However, Microsoft discontinued active support for AirSim in 2022 as development transitioned toward the newer Project AirSim platform. Consequently, the VECSR-A framework was also migrated to Project AirSim to maintain long-term compatibility and support.

We chose Unreal Engine 5 with Project AirSim to provide a high-fidelity simulation that better represents real world environments and complexity while maintaining deterministic reproducibility for evaluation. Unreal Engine is a real-time 3D development platform created by Epic Games for building video games, simulations, and interactive applications \cite{bib15}. Project AirSim is an advanced simulation platform that runs off of Unreal Engine 5. It was developed by Microsoft as an evolution of Microsoft AirSim for autonomous systems research and development. The platform is designed to support the simulation of drones, ground vehicles, and robotic systems in realistic virtual environments. Project AirSim provides tools for artificial intelligence training, sensor simulation, and autonomous navigation testing. The city environment used in this experiment was built using commercially available asset packs from the Unreal Marketplace.

Project AirSim provides several important advantages over the original AirSim platform. These improvements include compatibility with modern Unreal Engine versions, improved simulation scalability, enhanced rendering and environmental realism, and a more extensible architecture for future autonomous systems research. These capabilities enabled the development of highly realistic simulated environments for evaluating VECSR-A under a broader range of operational conditions. Additionally, Project AirSim improved integration flexibility for custom sensors, autonomous behaviors, and multi-agent experimentation, supporting more advanced testing and validation workflows.

\section{System Architecture}\label{sec3}

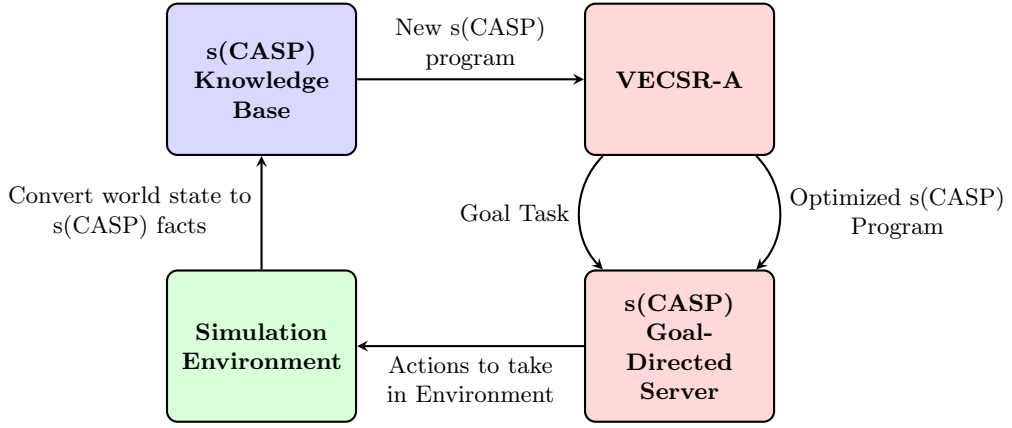
\begin{figure}[htbp]
\centering
\begin{tikzpicture}[auto, node distance=3cm]

\node[simblock] (airsim) {Simulation\\Environment};
\node[block, above=1.5cm of airsim] (kb) {s(CASP)\\Knowledge\\Base};
\node[serverblock, right= of kb] (vecsra) {VECSR-A};
\node[serverblock, below=1.5cm of vecsra] (server) {s(CASP)\\Goal-Directed\\Server};

\draw[arrow] (airsim) -- node[left, label] {Convert world state to\\s(CASP) facts} (kb);
\draw[arrow] (kb) -- node[above, label] {New s(CASP)\\program} (vecsra);

\draw[arrow, bend left=45] (vecsra) to node[right, label] {Optimized s(CASP)\\Program} (server);
\draw[arrow, bend right=45] (vecsra) to node[left, label] {Goal Task} (server);

\draw[arrow] (server) -- node[below, label] {Actions to take\\in Environment} (airsim);

\end{tikzpicture}%

\bigskip

\caption{VECSR-A architecture for commonsense reasoning.}%
\label{fig:architecture}%
\end{figure}%

We implement a fully realized version of VECSR-A by constructing a new architecture, supporting a variety of tasks based on the simulated environment. This architecture can be seen in Figure \ref{fig:architecture}. Our approach utilizes an s(CASP) knowledge base to represent UAV intelligence and rules, which we supplement with real-time perception facts derived from the environment. This synthesis generates a dynamic s(CASP) program for the VECSR-A agent. Upon receiving a goal task, the agent selects an action that is executed within the environment. The resulting state is then converted to facts and fed back into the knowledge base, closing the loop.

Our system, VECSR‑A, is goal‑driven: at the start of the simulation, the user specifies a high‑level task (e.g., `deliver a package') in a configuration file, and the agent autonomously goes to the relevant location and performs the task if possible. VECSR-A identifies task-relevant objects or locations from the simulation environment during top-down execution. It then uses state information and its knowledge base (e.g., `inspecting a building' is decomposed into the tasks of `get building location', `go to building', `take picture'), to bind variables to concrete objects in the current state. The top-down and query-driven execution mean rules unrelated to the current goal are ignored, allowing the system to focus on relevant information.

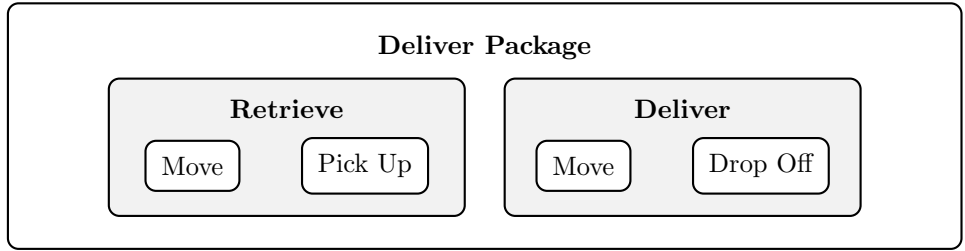
\begin{figure}[htbp]
\centering
\begin{tikzpicture}[
    outerbox/.style={rectangle, draw=black, thick, inner sep=12pt, rounded corners},
    innerbox/.style={rectangle, draw=black, thick, inner sep=8pt, rounded corners, fill=gray!10},
    leafbox/.style={rectangle, draw=black, thick, inner sep=6pt, rounded corners, fill=white},
    text centered
]

\node[outerbox] (root) {
    \begin{minipage}{0.9\textwidth}
        \centering
        \textbf{Deliver Package}
        
        \vspace{6pt}
        
        \begin{tikzpicture}
            \node[innerbox, minimum width=0.4\textwidth] (pickup) {
                \begin{minipage}{0.35\textwidth}
                    \centering
                    \textbf{Retrieve}
                    
                    \vspace{6pt}
                    
                    \begin{tikzpicture}
                        \node[leafbox, minimum width=0.15\textwidth] (goto1) {Move};
                        \node[leafbox, minimum width=0.15\textwidth, right=0.8cm of goto1] (Pick Up) {Pick Up};
                    \end{tikzpicture}
                \end{minipage}
            };
            
            \node[innerbox, minimum width=0.4\textwidth, right=0.5cm of pickup] (deliver) {
                \begin{minipage}{0.35\textwidth}
                    \centering
                    \textbf{Deliver}
                    
                    \vspace{6pt}
					
                    \begin{tikzpicture}
                        \node[leafbox, minimum width=0.15\textwidth] (goto2) {Move};
                        \node[leafbox, minimum width=0.15\textwidth, right=0.8cm of goto2] (Drop Off) {Drop Off};
                    \end{tikzpicture}
                \end{minipage}
            };
        \end{tikzpicture}
    \end{minipage}
    
};

\end{tikzpicture}%

\caption{Example hierarchical task decomposition.}%
\label{fig:hierarchical_nested}%
\vspace{-0.15in}
\end{figure}

The VECSR-A system decomposes a given goal task into a series of low-level controls that are implemented in the simulation. The overall completion of the task is done through multiple iterations of low-level action selections.  Figure \ref{fig:hierarchical_nested} demonstrates an example of this decomposition. The \texttt{deliver} goal task is divided into the four low-level tasks: \texttt{move}, \texttt{pickup}, and \texttt{drop off}. At each control cycle, the s(CASP) client sends an asynchronous query of the form: \texttt{?- choose\_action(Action, CurrentState, GoalState).} where \texttt{CurrentState} is an atom bound from the simulation state and \texttt{GoalState} represents the desired outcome. Using the knowledge base and current state, the s(CASP) server returns applicable atomic actions with justifications to reach the goal state, from which the system selects one for execution in Project AirSim. The chosen actions are first determined by fulfilled suggestion predicates, which determine a suitable action that moves the agent towards the goal state. The action must be legal, meaning it follows all constraints and is physically possible. Constraints over finite domains (e.g., distance or range queries) are handled natively during top-down execution without separate propagation passes \cite{bib1, bib3, bib5}. If no action is suggested, then the s(CASP) engine will instead select from all legal actions. 

This compositional design allows all tasks to build from the most basic level of primitive actions (e.g., \texttt{move}, \texttt{pick up}, \texttt{drop off}). Whether the drone is delivering a package, watering a plant, or inspecting a building, the overall action sequence is a composition of low-level actions. s(CASP) binds the specific sequence of actions at execution time based on the current task and state. This allows VECSR-A to run entirely new or unseen high-level tasks by decomposing these new tasks into a series of known lower-level actions.

\begin{lstlisting}[]
choose_action(Action, State1, State2) :- suggest(Action, State1, State2), legal_action(Action, State1).
choose_action(Action, State1, _) :- legal_action(Action, State1).
suggest(takeoff, _, _) :- is_landed(true).
...
legal_action(takeoff, _) :- is_landed(true).
...
\end{lstlisting}

By integrating goal-directed search, query-based execution, and native constraint handling, our system reasons over real-world states in near real-time \cite{bib3}. Because the agent computes only the information necessary to resolve immediate control queries rather than pre-computing all future states and possible groundings \cite{bib5}, each reasoning cycle takes approximately one second. VECSR-A uses the remaining time waiting for the drone to complete its physical action. Consequently, we achieve near real-time system performance constrained by how long it takes the drone to physically perform an action rather than computational overhead. We discuss runtime metrics in more detail in Section 5.1.

\section{Experimental Setup}\label{sec4}

\begin{figure*}[h]

\centering

\includegraphics[width = 300px]{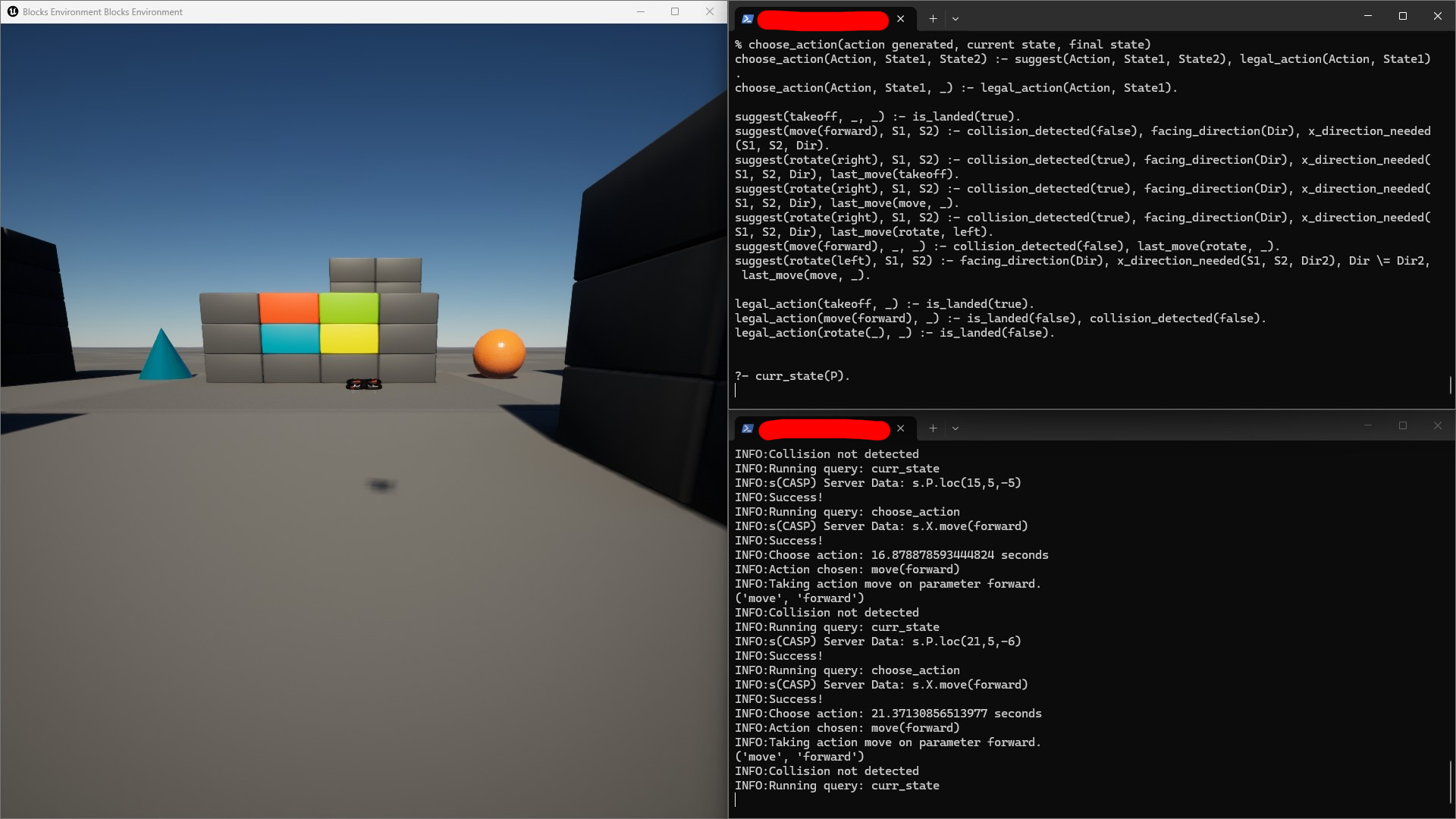}

\caption{Setup demonstrating all servers running VECSR-A locally}

\label{fig:vecsra_wsl}

\end{figure*}

We evaluate VECSR-A across three high-level capabilities:

\begin{description}

\item[\textbf{Autonomous Navigation and Search:}] This describes the agent's ability to logically determine a legal and efficient path from its current position to its target position. It also includes its ability to determine a target position based on its desired goal state. The agent needs to be able to determine target positions so that it may search for objects whose locations it may not immediately know, or to find objects that may have moved in a dynamic environment.

\medskip 
\item[\bf Environment Interaction:] Relates to an agent's ability to perform tasks involved with interacting with objects within its environment. The agent needs to figure out what actions are necessary to complete what tasks, execute on those actions, and retain information on how that action changed the environment.

\medskip 
\item[\bf Constraints and Adaptation:] In a highly dynamic environment, it's important that a drone agent is able to follow compliance settings. This includes physical constraints and legal or social constraints that the agent needs to follow while performing its tasks. In this experiment, we used spatial constraints (such as avoiding objects blocking the drone's movement), and specific task-based constraints (the agent shouldn't spray garden beds if a person is standing near the garden bed). 
\end{description}

\noindent These capabilities describe the overall metrics being used to evaluate the system. We perform five high-level tasks to do this evaluation:
\begin{description}

\item[\bf Task 1: Object Inspection:] The agent will move towards an unknown object, take a picture, and identify the object. If the picture taken isn't sufficient to identify the object, the agent will continue taking pictures until the object is properly identified. The perception aspect of this task is obtained directly form ground truth information in the simulation. This simulates the usage of machine learning perception tools with uncertainty predictions implemented over the object classifications.

\medskip 
\item[\bf Task 2: Debris Detection:] The agent will navigate around a building and identify objects within and around the building. If it detects debris on the building, then it will report that there is damage to the building. Otherwise, it will report that the building is safe. In this experiment, we simulate debris by placing undesired objects on the building (rocks/vehicles).

\medskip 
\item[\bf Task 3: Trespassers]: Similar to building debris, except the agent will instead explore a much larger section of the city and determine if certain vehicles have entered a no-driving zone. This task evaluates the drone's ability to identify objects and navigate the area to determine if there are trespassers within the area.

\medskip 
\item[\bf Task 4: Spraying:]The agent will move towards garden beds that have been targeted for spraying. The drone has to navigate to the desired garden bed and then interact with it by spraying it. It must then remember which garden beds it has sprayed and then avoid respraying them. The agent must also follow constraints, such as not spraying obstructed garden beds or not spraying when people are too close to the spraying area.

\medskip 
\item[\bf Task 5: Transport/Delivery:] There will be two points representing a pickup point and a drop-off point for an object delivery. The agent must navigate to the first point and interact by performing the pickup action. While it's holding the object, it must navigate to the second point and perform the drop-off action. 
\end{description}

\noindent 
It must be emphasized that all these tasks are performed by the drone autonomously using commonsense reasoning facilitated by s(CASP). 

\begin{figure}[h]
    \centering
    \includegraphics[width=0.7\textwidth]{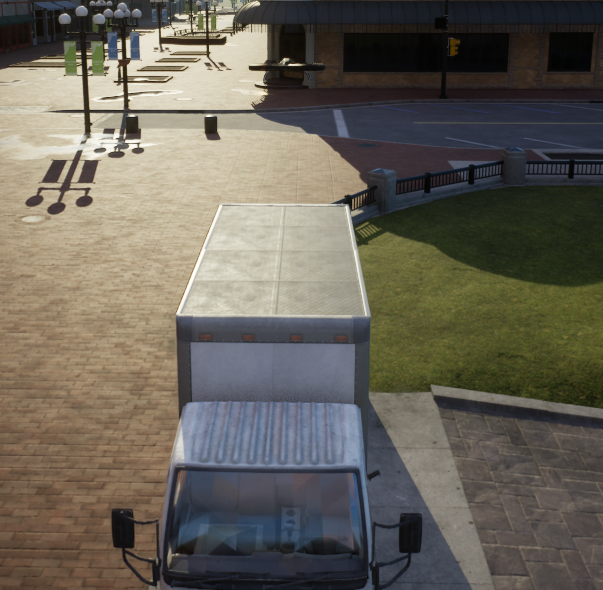}
    \caption{A delivery truck that is in a no driving zone.}
    \label{fig:trespasser}
\vspace{-0.15in}
\end{figure}

All experiments ran on a Windows 11 laptop (NVIDIA RTX 3050 Mobile GPU) on Unreal Engine 5 with Project AirSim. VECSR-A supports both onboard and server-based implementations. To demonstrate this, we used two Windows Subsystem for Linux (WSL) terminals on the same machine: one hosted the s(CASP) client (translating simulation state to logical facts and generating action queries), and the other hosted the s(CASP) server (performing constraint-based commonsense reasoning with the knowledge base and returning executable action sequences). Communication between Unreal Engine and the two WSL components used TCP sockets over localhost. VECSR-A then executes the cycle shown above to perform a series of actions in the Unreal environment. An example onboard setup is shown in Figure \ref{fig:vecsra_wsl}. The left side shows the active Unreal environment and the right side shows the s(CASP) host server and the Project Airsim client.

\section{Results}\label{sec5}
\begin{figure}[h]
    \centering
    \includegraphics[width=0.7\textwidth]{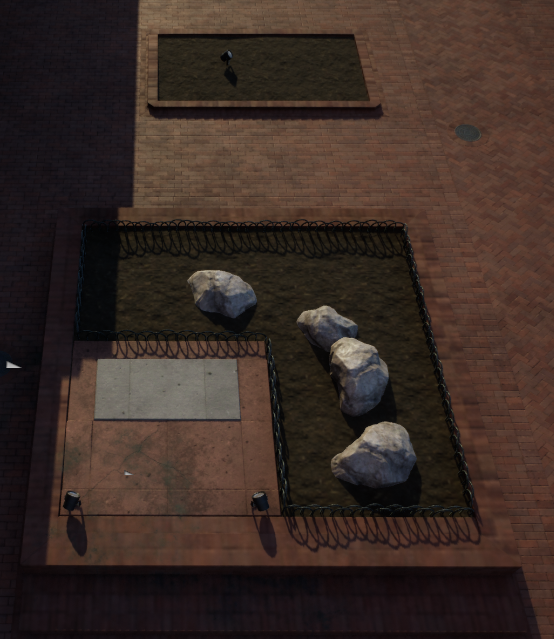}
    \caption{Two garden beds targeted for spraying in our Unreal city environment. One is obstructed by rubble and is unsuitable for spraying.}
    \label{fig:spraying}
\end{figure}

We next provide our results on the high-level tasks and evaluate our drone agent's performance in the context of  higher-level capabilities. We tested VECSR-A over five different scenarios that represent the five high-level tasks described earlier. VECSR-A successfully navigates the map and performs the five designated tasks accurately within the scenarios given to it. In the Unreal city map, we have VECSR-A identify unknown vehicles, move to a building to identify rocks and rubble, move around the map to find a vehicle in a no-driving zone, spray target garden beds, and pick up items from a garden bed and move them to a nearby delivery truck. We also observe that it effectively handles constraints on selected actions. This includes not moving to points with objects in the way, not spraying when there are obstructions, not spraying areas that have been sprayed already, only picking or dropping up an item when the drone is or isn't holding anything. Figure \ref{fig:spraying} shows the garden beds targeted for spraying, one which is obstructed by rubble. VECSR-A effectively determines that only the clear garden bed can be sprayed. 

VECSR-A, like its predecessor VECSR, is guaranteed to produce plans that are correct and executable. If a correct and executable plan cannot be produced, no plan is produced. In all the experiments performed, VECSR-A produced plans that achieved the goal state if it was possible to do so without violating constraints (i.e. spraying humans too close to flower beds, flying through objects, etc.). A video recording of the \texttt{deliver} task on the Project AirSim Simulation System can be viewed at the following URL: \url{https://drive.google.com/drive/folders/1LhpE5o9pKnRUxdQjrPokELGZSBWj6lrE?usp=drive_link.}.

\subsection{Runtime}\label{subsec1}
The following is an evaluation of the runtime for the VECSR-A agent. In Table \ref{table_runtime}, runtime is given in seconds. The `Total' column shows the total time to complete the task. The `Average Task Duration' column is the average time the drone takes to carry out the physical action including the reasoning step. The 's(CASP) time' column is the average time the logic engine takes to generate an action. For this runtime analysis, we ran VECSR-A on the same task multiple times to generate the average numbers used. 

\begin{table}[h]
\centering
\caption{VECSR-A Runtimes on the high-level tasks }\label{tab1}%
\begin{tabular}{@{}lccc@{}}
\toprule
Task & Total (sec) & Average Task Duration &s(CASP) Time \\
\midrule
Object Identification    &31.18    &10.39   &1.26   \\
Building Debris    &34.02    &8.505   &1.27   \\
Trespassers   &32.61    &10.87  &1.24   \\
Spraying   &22.52    &7.51   &1.32   \\
Transport/Delivery   &39.02    &7.804   &1.28   \\
\botrule
\label{table_runtime}
\end{tabular}
\vspace{-0.1in}
\end{table}

\subsection{Ablation Studies}\label{subsec2}
Here, we evaluate the ability of VECSR-A's individual components to fulfill the three high-level  capabilities. We separate various parts of the s(CASP) engine from the whole system to determine how effectively it performs these components.

\begin{description}

\item[\textbf{Autonomous Navigation and Search:}] We isolated the movement commands from the rest of the s(CASP) engine to determine how well it was able to navigate the environment. This component of the s(CASP) system was evaluated on ten different scenarios where it was given a target location to move to. The agent then simply needed to move to the location. The agent was properly able to get to the location in all scenarios, and was able to handle collisions when encountered.
\medskip

\item[\bf Environment Interaction:] Environment interaction is evaluated through the \texttt{suggested\_action} component of the s(CASP) system. This component determines what actions should be taken to complete a task based on the current state of the agent and the environment. It is therefore an effective measure of how well the agent is able to interact with the environment to complete its task. We separated this component and remove movement-based suggestions. We then evaluated it over ten different scenarios in which the agent only needed to figure out the next best environment interaction-based action. VECSR-A effectively determined the next required environment-based action it needed to take in the ten scenarios.

\medskip 
\item[\bf Constraints and Adaptation:] The \texttt{legal\_action} component of the s(CASP) system represents how well the system is able to determine actions that follow all necessary constraints. We evaluated over ten scenarios in which a list of ten actions was provided, and the system needed to determine how many of these actions were legal. VECSR-A correctly identified the legality of each of the ten actions for each scenario.

\end{description}

\subsection{Discussion}\label{sec6}

The VECSR-A framework demonstrates the viability of integrating high-fidelity simulation environments with symbolic commonsense reasoning to achieve robust UAV autonomy. As stated above, our approach uses Answer Set Programming (ASP) to reason under incomplete information, employing assumption-based reasoning where the absence of evidence is treated as a lack of constraint \cite{bib3}. Our agent leverages s(CASP) to mimic human-like piloting logic, such as dynamically avoiding spraying garden beds when pedestrians are nearby or bypassing obstructed paths during debris detection tasks.

\subsubsection{Advantages}
A critical advantage over traditional ASP solvers is s(CASP)'s goal-directed execution. By avoiding exhaustive grounding of the entire state space, our system achieves near real-time decision-making, which is essential for closed-loop control in dynamic environments.

The modularity of the knowledge base represents a significant departure from data-driven approaches. Unlike neural networks that require extensive retraining to adapt to new constraints \cite{bib8}, VECSR-A allows both task decomposition and mission parameter updates by adding or modifying rules within the s(CASP) program. For instance, in our transport and spraying experiments, the agent adheres to new spatial constraints (e.g., ``do not spray if a person is present") without architectural changes. This top-down reasoning evaluates constraints only when relevant to the current query, enabling adaptation to changing mission environments \cite{bib1}.

Our work contributes to Explainable AI by providing a verifiable justification tree for every action taken by the UAV agent \cite{bib3}. While deep learning models often function as ``black boxes" where failure modes are opaque, VECSR-A generates explicit logical proofs for its decisions. This transparency is crucial for mission-critical operations, allowing human operators to understand exactly why a specific action was suggested or skipped.

This aligns with recent calls for interpretable logic over black-box Machine Learning models. It builds upon efforts that combine learning-based perception with symbolic constraint satisfaction \cite{bib17,bib18,bib8, bib9}. Furthermore, the high-fidelity simulation provided by Project AirSim on Unreal Engine 5 ensures that the logical reasoning operates within a human-based and physically realistic context, leveraging advances in simulation modeling established in prior autonomous navigation research \cite{bib11}.

\subsubsection{Future Work}
The one-second reasoning cycle observed in our experiments (Table \ref{table_runtime}) is well-suited for deliberative tasks where the environment remains relatively static between decisions. However, this latency may be prohibitive for highly reactive scenarios that require millisecond-level responses to rapidly moving threats, such as actively falling debris or fast-moving vehicles. Additionally, our current implementation assumes perfect perception; future iterations could couple VECSR-A with probabilistic action models to handle sensor noise and uncertain outcomes more effectively. It is also possible to implement uncertainty modeling and consistency checking in the s(CASP) model to alleviate the impact of incorrect perceptions of the environment \cite{bib16}.

Future work could also build on this framework with multi-agent scenarios. By enabling multiple UAVs to act on a unified s(CASP) knowledge base for distributed constraint solving. For example, in agricultural applications, multiple drones could coordinate to avoid spraying the same crop area, simultaneously or collaboratively narrow down search areas and limit backtracking \cite{bib6}.

\section{Conclusion}\label{sec7}

In this paper, we fully implemented the VECSR-A framework to support complex, multi-step autonomous task execution using s(CASP) for commonsense reasoning within a simulated Unreal Engine 5 city environment. By integrating Project AirSim with goal-directed answer set programming, we demonstrated a system capable of performing hierarchical tasks (Autonomous Navigation and Search, Environment Interaction, and Constraints and Adaption) while adhering to dynamic environmental constraints. Our simulation leverages the high-fidelity visual capabilities established in prior works ~\cite{bib1, bib2, bib5} to provide a realistic testbed for symbolic reasoning.

Our experimental results confirm that VECSR-A successfully fulfills mission objectives in near real-time while dynamically satisfying complex constraints and navigating environmental obstacles. The goal-directed nature of s(CASP) facilitates near real-time decision-making, with the logic engine requiring approximately one second per cycle to generate actionable plans. Crucially, the system exhibits robust adaptability; it autonomously generates corrective actions, such as bypassing obstructed garden beds or re-evaluating paths around debris, without requiring manual intervention or model retraining. This flexibility highlights the efficiency of top-down reasoning over traditional grounding methods in dynamic settings.

We demonstrated the utility of symbolic reasoning in achieving explainable autonomy. In contrast to black-box deep learning models, which often obscure the rationale behind specific failure modes, VECSR-A generates a verifiable justification tree for every action \cite{bib5}. This capability ensures that mission-critical decisions are both logically sound and inherently interpretable, fostering greater transparency and trust for human operators in safety-critical scenarios.

VECSR-A provides a scalable foundation for autonomous systems where reliability and explainability are paramount. As UAVs become increasingly deployed in disaster response, inspection, and agriculture, the ability to reason over complex constraints using human-like commonsense logic will be essential for building trust and ensuring mission success in dynamic, unpredictable environments.

\bibliography{sn-bibliography}

\end{document}